\documentstyle[prb,aps,preprint,epsf]{revtex}
\def\geqap{\,\raise 2pt \hbox{$>\kern-11pt \lower 5pt \hbox{$\sim$}$}\,}
\def\leqap{\,\raise 2pt \hbox{$<\kern-10pt \lower 5pt \hbox{$\sim$}$}\,}
\begin{document}
\draft
\title{Jahn-Teller effect  and Electron Correlation in manganites}
\author{Ryo Maezono}
\address{Theory of Condensed Matter Group, Cavendish Laboratory, 
University of Cambridge, \\
Madingley Road, Cambridge CB3 0HE, U.K.}
\address{National Institute for Materials Science,
Computational Materials Science Division, \\
Sengen 1-2-1, Tsukuba, Ibaragi, 305-0047, Japan.}
\author{Naoto Nagaosa}
\address{Department of Applied Physics, University of Tokyo,
Bunkyo-ku, Tokyo 113-8656, Japan.}
\address{Correlated Electron Research Center (CERC), 
National Institute of Advanced Industrial Science and Technology (AIST), 
Tsukuba Central 4, Tsukuba 305-8562, Japan.}

\date{\today}
\maketitle
%
\begin{abstract}
Jahn-Teller (JT) effect both in the absence and presence of 
the strong Coulomb correlation is studied theoretically focusing 
on the reduction $\Delta K$ of the kinetic energy gain which is directly 
related to the spin wave stiffness.
Without the Coulomb interaction, the perturbative analysis 
gives $\Delta K / (g^2/M\Omega^2) \cong 0.05-0.13$
depending on the electron number
[$g$: electron-phonon(el-ph) coupling constant, $M$: mass of the oxygen atom, 
 $\Omega$: frequency of the phonon].
Although there occurs many channels of the JT el-ph interaction
in the multiband system, the final results of $\Delta K$
roughly scales with the density of states at the Fermi energy.
In the limit of strong electron correlation, the magnitude of the orbital 
polarization saturate and the relevant degrees of freedom are the 
direction (phase) of it. An effective action is derived for the 
phase variable including the effect of the JT interaction.
In this limit, JT interaction is {\it{ enhanced}} compared with the 
non-interacting case, and $\Delta K$ is given by the lattice relaxation 
energy $E_{L}$ for the localized electrons, although the electrons remains
itinerant.
Discussion on experiments are given based on these theoretical results.
\end{abstract}
%
%
\narrowtext
%
\section{Introduction}
%
It has been recognized that in the strongly correlated electronic systems 
both the electron-phonon (el-ph) and electron-electron (el-el) interaction 
are enhanced and play important roles.
In manganites, the colossal magneto-resistance (CMR) 
\cite{CHA93,HEL93,TOK95,JIN94} has been discussed from 
both points of view.
In this system the ferromagnetism is basically explained by the double 
exchange model.\cite{GEN51,AND55,DEG60}
However the orbital degeneracy of the $e_g$-states is considered to be very 
important, and the orbital ordering or disordering is the crucial issue for 
the understanding of the CMR effect.
\cite{ISH97,NAG98,KIL98,END99,KHA00,TAK00,MAE00c,BRK01}
Therefore there are two viewpoints on this problem, namely the orbital 
degrees of freedom is governed by (i) the Jahn-Teller (JT) electron-phonon 
coupling or (ii) the electron correlation. 
In the former case, the change of the bandwidth due to the crossover 
from small to large JT-polaron is the key mechanism of the CMR effect, 
and the JT effect is assumed to be negligible in the ferromagnetic 
metallic state. \cite{MIL95,MIL96a,MIL96b,MIL96c,ROD96}
This picture appear contradicting  with the orbital orderings, which 
require rather strong coupling,  surrounding the ferromagnetic region
in the phase diagram.
\cite{END99,KAJ99,MRT98a,TOK00,AKI99,KUB00,MIT00,MAE98a,MAE99b,BRK98,SHE99,OKA00a,OKA00b,FAN00}
However it might be the case that the metallic screening weakens 
the el-ph interaction and/or the el-el interaction, 
and only in the ferromagnetic metallic state 
both of them could be neglected although the Hund's coupling is strong 
enough to polarize the spins perfectly.  
On the other hand, it has been recognized by several authors 
\cite{ISH97,NAG98,KIL98,END99,KHA00,TAK00,MAE00c,BRK01,MAE98a,OKA00}
that the Coulomb interaction play the important role in the physics 
of the orbital degrees of freedom, and the JT-interaction is the 
secondary effect.
\par
%
In this paper we revisited this issue by considering both the el-ph 
and el-el interaction, because both of them are considered to be 
relevant. Then the interplay between these two interactions
is the key issue.  
As shown below, the metallic screening of the JT el-ph interaction 
does not occur in contrast to the coupling between the breathing 
mode and the charge fluctuation.  
Coulomb interaction and JT effect collaborate with each other, 
and it is concluded that JT effect is {\it{enhanced}} by the 
el-el interaction by comparing the two limits of zero and strong 
electron correlations. 
Similar idea has been proposed by one of the authors \cite{NAG98}
in the context of the large-$d$ approximation  
\cite{MIL95,MIL96a,MIL96b,MIL96c,ROD96,BEN98}, 
where the single site model embedded in the dymanical 
environment is considered \cite{GEO96}. 
To our knowledge the effect of the JT phonon scattering with $d=3$ has 
not yet been exploited so much, and we found that the  
el-ph coupling which modulates the transfer integral 
becomes relevant in both limits of the non-interacting and the strong 
correlation limit.
Another important issue is how the el-ph intereaction manifests itself 
in the spin stiffness observed in neutron scattering experiments 
\cite{MAE99a}. 
Because of the half-metallicity, the stiffness is roughly proportional 
to the kinetic energy.\cite{QUI98}
The polaronic effect on the kinetic energy is therefore expected to 
appear as a reduction of the stiffness. \cite{MAE99a,QUI98}. 
The apparent absence of this reduction has led to the conclusion that 
el-ph interaction is irrelevant in the ferromagnetic metallic state. 
However we found that JT el-ph interaction is there even in the 
ferromagnetic state although it might be hidden in the inaccuracy of 
the theoretical estimation of the bare spin stiffness compared with 
the experimental ones. 
We expect near 3$\%$ of the reduction of the kinetic energy and spin 
stiffness due to the JT interaction as a lower bound estimated 
in the non-interacting limit.
\par
We first consider the non-interacting electrons with orbital degeneracy 
with JT el-ph coupling. \cite{MAE98a} 
The single-band  model where the
charge density is coupled to the phonon  can be treated 
by means of the canonical transformation, leading to the
reduction of the kinetic energy via the Debye-Waller factor. 
\cite{MAH90} 
This argument is applicable to the interaction with the breathing mode. 
However this method can not be generalized to the JT el-ph interaction 
with multiband electronic structure 
because it includes off-diagonal components with respect to orbital 
indices. 
Due to this difficulty, it is hard to apply the same argument as 
the breathing mode case to clarify whether the JT polaron also leads to 
the reduction of the kinetic energy or not. 
Therefore we employ the perturbative analysis on the JT el-ph coupling 
to estimate the reduction of the kinetic energy and spin stiffness.   
Although many channels contribute, each of which can be even negative,  
the resultant reduction in the kinetic energy gain
is roughly proportional to the density of states 
at the Fermi energy, and has a peak at around $n=0.6$ and $n=1.4$, 
where $n$ denotes the filling of the $e_g$ band.
\par
Next we consider the strong correlation limit by employing the effective 
Lagrangian which is derived as a projection onto the polarized orbital 
state. \cite{MAE98a} 
In this limit, the magnitude of the orbital polarization has been 
saturated, and the only degrees of freedom coupled to the JT phonon is 
therefore the direction of the polarization (which corresponds to the shape 
of the orbital ordering). 
The direction is determined by the double exchange interaction,
which is of the order of the transfer integral, $t_0$.
The quantum fluctuation of the direction (physically the fluctuation
of the orbital shape) is couples to the lattice deformation via the
el-ph interaction, leading to the lattice relaxation
energy $E_L=g^2/M\Omega^2$ ($g$, $M$ and $\Omega$ denote the JT el-ph 
coupling constant, the atomic mass, and the phonon frequency, respectively).
The characteristic frequency  $\omega_n$ of the orbital fluctuation
is of the order of the transfer integral $t_0$, and 
is larger than the phonon frequency $\Omega$.
Namely the orbital deformation can follow up 
the phonon, leading to the kinetic energy correction, $\Delta K \sim E_L$.
The phonon frequency ($\Omega$-) dependence of the kinetic energy 
correction in this case, $\left[\Delta K_{U\rightarrow\infty}
\left(\Omega\right)/E_L\right]\sim O\left(1\right)$, 
is different from that of non-interacting electrons case, 
$\left[\Delta K_{U=0}\left(\Omega\right)/E_L\right]\sim 
\left[\left(\Omega/t_0\right)/\left(1+\Omega/t_0\right)^2\right]$.
Thus the strong correlation enhances the JT effect, 
in sharp contrast to the case of breathing mode where the Coulomb 
interaction reduces the el-ph interaction. 
This is understood rather easily.
In the case of the breathing mode, the Coulomb interaction suppresses 
the charge fluctuation while the breathing mode induces it, i.e.,
these two interactions compete with each other, and the former suppresses 
the latter. 
Furthermore the metallic screening effect also  suppresses the charge 
fluctuation (This situation has been discussed in the context of the 
vertex correction 
of the el-ph interaction in the physics of high-$T_c$ 
cuprates \cite{shen}).
On the other hand, in the JT mode case, where el-el and el-ph interactions  
collaborate to induce the orbital 
pseudospin moment, the former enhances the latter and vice versa. 
Therefore there is no reason to expect the weakening of the JT el-ph 
interaction with the doping when the strong el-el interaction keeps 
the orbital pseudo-spin moment to saturate even in the ferromagneitc 
metallic state. 
\par
The plan of this paper follows. The perturbative analysis of the JT 
el-ph interaction for the noninteracting electrons with orbital 
degeneracy is given in Section II.  
The strong correlation limit is studied in Sec. III, and discussion 
and conclusions are presented in Sec. IV.
%
\section{non-interacting limit}
\begin{figure}[p]
\begin{center}
\vspace{0mm}
\hspace{0mm}
\epsfxsize=15cm
\epsfbox{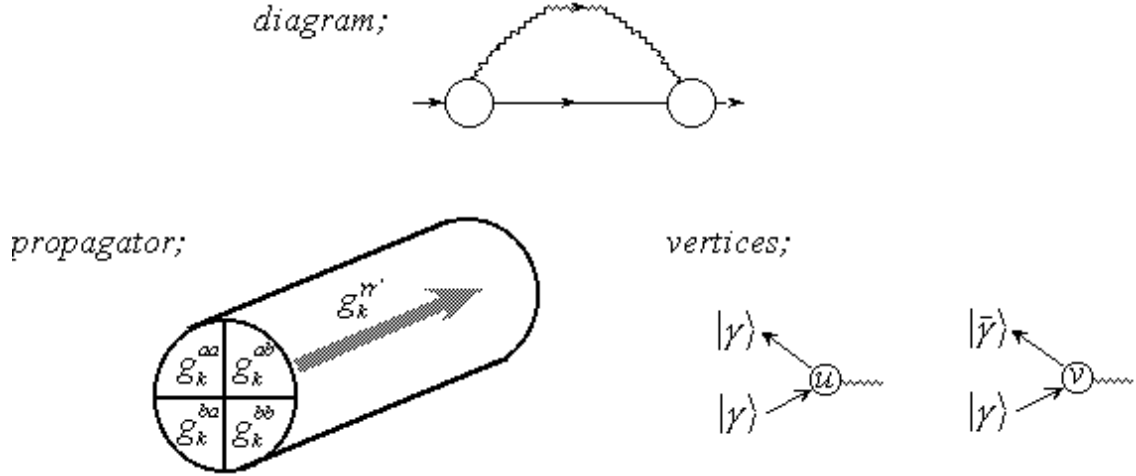}
\vspace{0mm}
\caption[aaa]{Diagrams of the self-energy due to the JT interaction.
Correponding to the doubly degenerate orbitals, there are
two kinds of vertices and the diagram is composed with four-channeled 
propagator.}
\label{JF1}
\end{center}
\end{figure}

\begin{figure}[p]
\begin{center}
\vspace{0mm}
\hspace{0mm}
\epsfxsize=15cm
\epsfbox{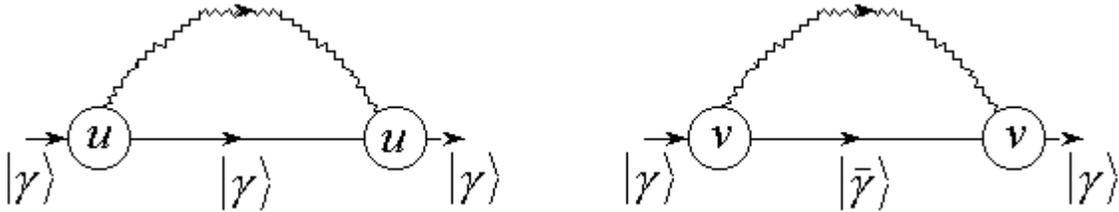}
\vspace{0mm}
\caption[aaa]{Corresponding diagrams of the terms in Eq. (11).}
\label{JF2}
\end{center}
\end{figure}

%
The JT interaction is given as
%
\begin{equation}
\label{Eq.J1}
H_{JT}=g\sum\limits_j {\left[ {\left({d_{ja}^{\dagger}d_{ja}-
d_{jb}^{\dagger}d_{jb}}\right)\cdot Q_{u,j}
+\left({d_{ja}^{\dagger}d_{jb}+d_{jb}^{\dagger}d_{ja}} \right)
\cdot Q_{v,j}} \right]} \ ,
\end{equation}
%
with a coupling constant $g$.
The spinless operator for the half-metallic ferromagnetic phase, 
$d_{j\gamma }^{\dagger}$, creates a spin polarized 
$e_g$ electron with orbital $\gamma$ [$=a(d_{x^2-y^2}),b(d_{3z^2-r^2})$]
at site $j$.
$Q_u$ and $Q_v$ denote the normal coordinates of the displacement
of the oxygen ions ${\Delta_\alpha}$ ($\alpha=x,y,z$):
$Q_u=\left({2\Delta _z-\Delta _x-\Delta _y}\right) / {\sqrt 6}$, 
$Q_v=\left({\Delta _x-\Delta _y}\right)/{\sqrt 2}$.
Let us consider the kinetic energy correction due to the JT phonon
scattering with a two-band model,
%
\begin{equation}
\label{Eq.J2}
H=\sum\limits_{i\delta,\gamma \gamma '} {t_{i,i+\delta }^{\gamma \gamma'}
\cdot d_{i\gamma}^{\dagger}d_{i+\delta,\gamma'}}
+\sum\limits_j {\left[ {{1 \over {2M}}\vec P_j\cdot \vec P_j+{{M\Omega ^2} 
\over 2}\vec Q_j\cdot \vec Q_j} \right]}+H_{JT} \ .
\end{equation}
%
$\left\{t_{i,i+\delta }^{\gamma \gamma'}\right\}$ are
realistic anisotropic hopping intensities given in ref.\cite{MAE98a}.
$\vec Q_j$ is defined as $\left(Q_{u,j},Q_{v,j}\right)^t$.
$M$, $\Omega$, and $\vec P_j$ denote the atomic mass, the elastic constant,
and the canonical momentum of $\vec Q_j$, respectively.
For a simplified model without orbital indices, the canonical transformation 
is a standard method to deal with the electron-phonon interaction, leading to
the kinetic energy reduction by the Debye-Waller factor. \cite{MAH90}
The canonical transformation for the JT phonon has off-diagonal elements
with respect to orbital indices, due to which the application
of this method becomes complicated.
We therefore employ the perturbative calculation  of the kinetic
energy $K$ up to the second order with respect to the coupling constant $g$, 
as
%
\begin{equation}
\label{Eq.J3}
K=K_0+\Delta K \ ,\  \Delta K \sim O\left(g^2\right) \ .
\end{equation}
%
$\Delta K$ is expressed by diagrams of the self-energy
shown in Fig. \ref{JF1}. 
%
\noindent
%
There are two vertices corresponding to $Q_u$- and $Q_v$-scatterings.
The propagator takes $2\times 2$-matrix form with respect to the orbital
indices.
Annihilation and creation operators of the JT phonons
are introduced as
%
\begin{equation}
\label{Eq.J4}
Q_{u,v}\left( t \right)=\sqrt {{1 \over {2M\Omega }}}\cdot 
\left( {a_{u,v}\cdot e^{-i\Omega t}+a_{u,v}^{\dagger}\cdot 
e^{i\Omega t}} \right)\ .
\end{equation}
%
With the momentum representation of the operators $c=\left(a,d\right)$,
%
\begin{equation}
\label{Eq.J5}
c_j\left( \tau  \right)={1 \over {\sqrt {\beta N}}}\sum\limits_{j,l} 
{c_q\left( {i\omega _l} \right)\cdot e^{iqR_j-i\omega _l\tau }} \ \ ,\ \ 
i\omega_l \rightarrow z \ , 
\end{equation}
%
with Matsubara frequency $i\omega_l$, the propagators of electrons and 
phonons are given as
%
\begin{eqnarray}
\label{Eq.J6}
g_k^{\gamma\gamma'} \left( z \right) 
&=&  
- T\left\langle {d_{k\gamma}\left(z\right)d_{k\gamma'}^\dagger\left(z\right)} 
\right\rangle_{0;i\omega_l=z}=\left[ {\left( {z + \mu } \right)\delta _{\gamma \gamma'} - \varepsilon _k^{\gamma \gamma '} } \right]_{\gamma \gamma '}^{ - 1}  = \frac{{A_{ + ;k}^{\gamma \gamma '} }}{{z - \Xi _k^{\left(  +  \right)} }} + \frac{{A_{ - ;k}^{\gamma \gamma '} }}{{z - \Xi _k^{\left(  -  \right)} }}
\ ,
\\
\nonumber \\
D_q^{u\left( v \right)}\left( z \right)&=&
{1 \over {2NM\Omega }}\cdot \left[ {T\left\langle {a_{q,u\left( v \right)}
\left( z \right)a_{q,u\left( v \right)}^{\dagger}\left( z \right)} 
\right\rangle+T\left\langle {a_{-q,u\left( v \right)}^{\dagger}
\left( z \right)
a_{-q,u\left( v \right)}^{}\left( z \right)} \right\rangle} \right]
\nonumber \\
&=&{1 \over {2NM\Omega }}\cdot \left[ {{1 \over {z+\Omega }}-{1 \over 
{z-\Omega }}} \right]=D_q^{}\left( z \right) \ ,
\end{eqnarray}
%
respectively.
The $u$- and $v$-modes have the same mass and frequency because they belong
to the same irreducible representation, and then the same phonon propagator.
Here we neglected the inter-cluster coupling of the vibration and hence 
the wavelength dependence of the phonon frequency.
Coefficients $A^{\gamma\gamma'}_{\pm;k}$ in Eq. (\ref{Eq.J6}) are defined as
%
\begin{eqnarray}
\label{Eq.J8}
A_{ + ;k}^{\gamma \gamma }=\frac{{\Xi _k^{\left(  +  \right)}  - \xi _k^{\bar \gamma } }}{{\Xi _k^{\left(  +  \right)}  - \Xi _k^{\left(  -  \right)} }}
\quad , \quad
A_{ - ;k}^{\gamma \gamma }  =  - \frac{{\Xi _k^{\left(  -  \right)}  - \xi _k^{\bar \gamma } }}{{\Xi _k^{\left(  +  \right)}  - \Xi _k^{\left(  -  \right)} }}
\nonumber\  , 
\\
A_{ + ;k}^{\gamma \bar \gamma }  = \frac{{ - \varepsilon _k^{ab} }}{{\Xi _k^{\left(  +  \right)}  - \Xi _k^{\left(  -  \right)} }}
\quad , \quad
A_{ - ;k}^{\gamma \bar \gamma}  =  - \frac{{ - \varepsilon _k^{ab} }}{{\Xi _k^{\left(  +  \right)}  - \Xi _k^{\left(  -  \right)} }}\ ,
\end{eqnarray}
%
with dispersion relations of the hybridized bands given as
%
\begin{equation}
\label{Eq.J9}
\Xi _k^{\left( \pm  \right)}
={1 \over 2}\left[{\left( {\xi _k^a+\xi _k^b} \right)\pm 
\sqrt{\left({\xi _k^a-\xi _k^b}\right)^2+4\left(\varepsilon_k^{ab}\right)^2
}} \right]
\quad ,\quad \xi _k^\gamma =\varepsilon _k^\gamma -\mu \ .
\end{equation}
%
The orbital index $\bar\gamma$ is used as $\bar a=b$ and $\bar b=a$.
$\varepsilon _k^{aa}\left( {=\varepsilon _k^a} \right)$,
$\varepsilon _k^{bb}\left( {=\varepsilon _k^b} \right)$, and
$\varepsilon _k^{ab}$ are the cosine dispersions with overlap
integrals between the orbitals 
$\left| a \right\rangle =\left| {x^2-y^2} \right\rangle $ and
$\left| b \right\rangle =\left| {3z^2-r^2} \right\rangle $.
With the propagators, the kinetic energy correction $\Delta K$ is given as
%
\begin{eqnarray}
\label{Eq.J10}
\Delta K&=&{1 \over 2}\sum\limits_{l}\sum\limits_{k,\gamma\gamma'} 
{\sum\limits_{\gamma _1\gamma_2} {\varepsilon _k^{\gamma \gamma '}
g_{\vec k}^{\gamma \gamma _1}\left( {i\omega _l} \right)\cdot 
\Sigma _{\vec k}^{\gamma _1\gamma_2}\left( {i\omega _l} \right)\cdot 
g_{\vec k}^{\gamma _2\gamma '}\left( {i\omega _l} \right)}}
\ , \\ 
\Sigma_{\vec k}^{\gamma_1\gamma_2}\left( {i\omega_l} \right)&=&
-{{g^2} \over \beta }\sum\limits_q {\oint_c {{{dz} \over {2\pi i}}}
\!\cdot\! f\left( z \right)\left[ {g_{\vec k-\vec q}^{\gamma _1\gamma _2}
\left( z \right)+g_{\vec k-\vec q}^{\bar \gamma _1\bar \gamma _2}
\left( z \right)} \right]}\!\cdot\! D_q^{}\left( {i\omega _l-z} \right) \ ,
\end{eqnarray}
%
where the contour $c$ surrounds the poles of the fermi distribution 
function $f\left(z\right)$.
In Eq. (11), ${g_{\vec k-\vec q}^{\gamma _1\gamma _2}
\left( z \right)}$ and ${g_{\vec k-\vec q}^{\bar \gamma _1\bar \gamma _2}
\left( z \right)}$ correspond to the scattering by $u$- and $v$-vertex,
respectively. 
These contributions are represented by diagrams shown in Fig. \ref{JF2}.
%
\noindent
%
Introducing a notation,
%
\begin{equation}
\label{Eq.J12}
I_{\left( {\gamma _2 \gamma '_2 ;\gamma _3 \gamma '_3 } \right)}^{\left( {\gamma _1 \gamma '_1 } \right)} \left( k \right) = \sum\limits_l {\Sigma _{\vec k}^{\gamma _1 \gamma '_1 } \left( {i\omega _l } \right) \cdot g_{\vec k}^{\gamma _2 \gamma '_2 } \left( {i\omega _l } \right)g_{\vec k}^{\gamma _3 \gamma '_3 } \left( {i\omega _l } \right)} \ ,
\end{equation}
%
Eq. (\ref{Eq.J10}) is expanded as,
%
\begin{eqnarray}
\label{Eq.J13}
\Delta K 
&=& 
\frac{1}{2}\sum\limits_k {\left[ {\varepsilon _k^{aa}  \cdot I_{\left( {aa;aa} \right)}^{\left( {aa} \right)}  + \varepsilon _k^{aa}  \cdot I_{\left( {ab;ab} \right)}^{\left( {aa} \right)}  + 2\varepsilon _k^{ab}  \cdot I_{\left( {ab;aa} \right)}^{\left( {aa} \right)} } \right]} 
\nonumber\\
&&
 + \frac{1}{2}\sum\limits_k {\left[ 
{\varepsilon_k^{bb}\cdot I_{\left({bb;bb} \right)}^{\left( {bb}\right)}  +
 \varepsilon_k^{bb}\cdot I_{\left({ab;ab} \right)}^{\left( {bb}\right)}  + 
2\varepsilon_k^{ab}\cdot I_{\left({ab;bb} \right)}^{\left( {aa}\right)}} 
\right]} 
\nonumber\\
&&
 + \sum\limits_k {\left[ {\varepsilon _k^{aa}  \cdot I_{\left( {ab;aa} \right)}^{\left( {ab} \right)}  + \varepsilon _k^{bb}  \cdot I_{\left( {ab;bb} \right)}^{\left( {ab} \right)}  + \varepsilon _k^{ab}  \cdot \left( {I_{\left( {ab;ab} \right)}^{\left( {ab} \right)}  + I_{\left( {aa;bb} \right)}^{\left( {ab} \right)} } \right)} \right]} \ .
\end{eqnarray}
%
Three terms correspond to the 
contribution from $\Sigma^{aa}_{k}$,
$\Sigma^{bb}_{k}\left(=\Sigma^{aa}_{k}\right)$, and
$\Sigma^{ab}_{k}$, respectively.
Note that the upper (lower) suffix $\gamma_n\gamma'_n $ of 
$I_{\left( {\gamma _2 \gamma '_2 ;\gamma _3 \gamma '_3 } \right)}
^{\left( {\gamma _1 \gamma '_1 } \right)}$
means that the correpsonding contribution comes from
a diagram composed of a propagator $g^{\gamma_n\gamma'_n}_{k-q}
\ \left(g^{\gamma_n\gamma'_n}_{k}\right)$ for the state $\left|k-q\right
\rangle \ \left(\left|k\right
\rangle\right)$
[When one finds $\left(ab\right)$ in the upper (lower) suffix, that 
contribution contains the hybridization during the 
propagation with the wave vector $\left|k-q\right \rangle \ \left(\left|k
\right\rangle\right)$].
In Appendix A are given concrete forms of 
$I_{\left( {\gamma _2 \gamma '_2 ;\gamma _3 \gamma '_3 } 
\right)}^{\left( {\gamma _1 \gamma '_1 } \right)}$.
\par
\begin{figure}[p]
\begin{center}
\vspace{0mm}
\hspace{0mm}
\epsfxsize=17cm
\epsfbox{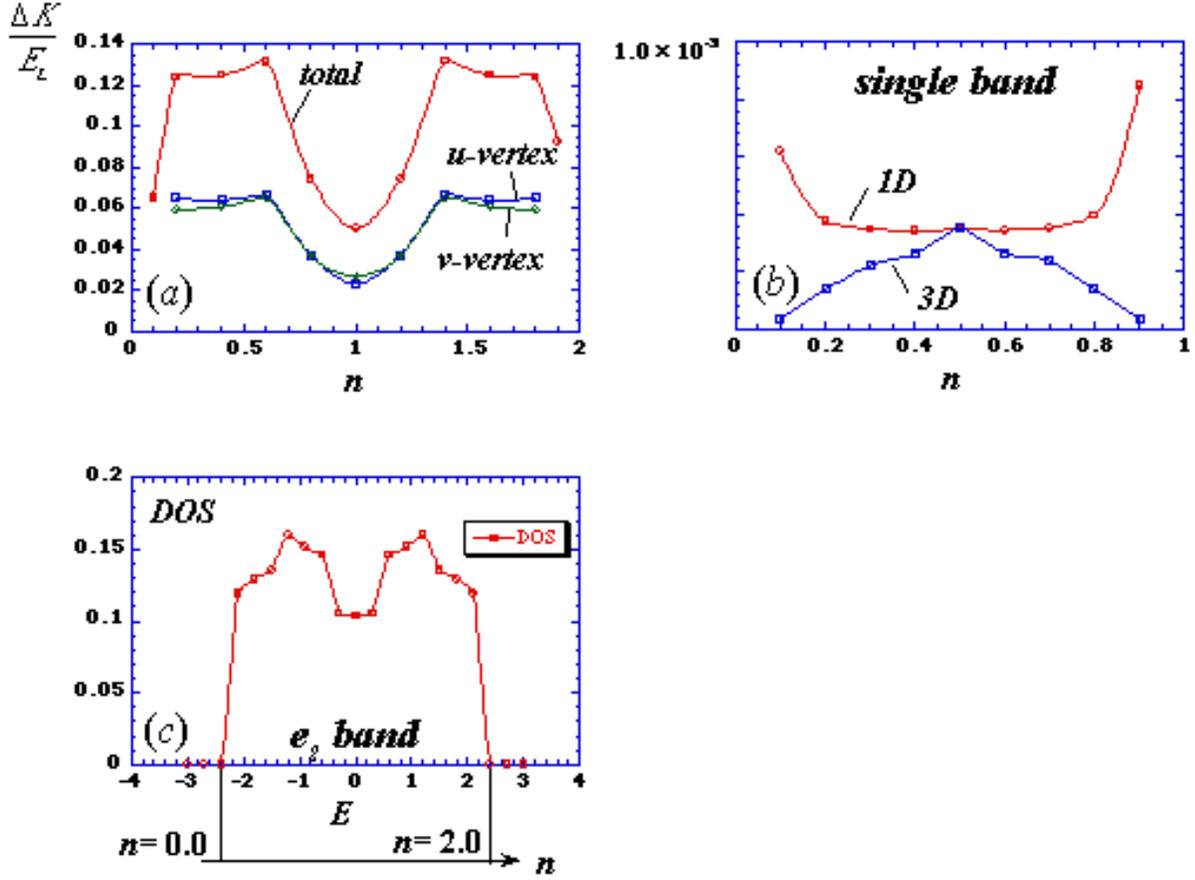}
\vspace{0mm}
\caption[aaa]{Kinetic energy reduction as a function of the filling $n$ 
for non-interacting electrons.
(a) Total result and partial contributions from $u$- and $v$-vertices in a 
realistic two-band model with  $e_g$ anisotropy.
(b) Result for one and three dimensional single band cases with breathing 
type phonons.
(c) Density of states as a function of the fermi energy for given $n$ 
    (lower $n$-axis) in the system used for (a).}
\label{JF3}
\end{center}
\end{figure}

%
Numerical results are shown in Fig. \ref{JF3} with parameters $t_0=0.72$ eV
\cite{MAE98a} and $\Omega=0.05$ eV.\cite{YMM00}
The kinetic reduction $\Delta K$ is calculated as a function of the 
filling $n$.
%
\noindent
%
Panel (a) shows the total result [Eq. (\ref{Eq.J13})] and partial 
contributions due to the $u$- and $v$-vertices.
The particle-hole symmetry with respect to the axis $n=1.0$ is seen.
The positive definite result is obtained for the whole range of $n$.
In order to understand the origin of the $n$-dependence in the plot (a), 
we also calculate the simpler case with breathing type phonons
and single band electrons [Eq. (\ref{Eq.JA8})] in one and three dimensions, 
as shown in the panel (b).
In this case $\Delta K/E_L$ roughly scales to
the density of states at the fermi level $N(\varepsilon_F)$
for given $n$ (see Appendix B).
According with this expectation, the result has the minimum (maximum) 
at $n=0.5$ for one (three) dimensional case.
The positive definite result in the plot (b) is consistent with the 
consequence from the canonical transformation method. \cite{MAH90}
Though the expression for the realistic doubly-degenerate $e_g$ case 
[Eq. (\ref{Eq.J13}) and the panel (a)] is much more complicated, the result 
also seems to scale to the density of states.
For the comparison, the plot of the density of states in this case
is shown in the panel (c).
Values of $n$ giving the peak and the dip in panel (a) and (c)
actually coincide each other.
As discussed in Appendix B,  $k$- (wave vector) points near the 
Fermi level contribute dominantly to $\Delta K$.
Unless the $n$-dependence of each contributing value is so sensitive,
$\Delta K$ simply scales to the population of the contributing 
$k$-points and hence $N(\varepsilon_F\left(n\right))$ for given $n$.
This gives a rough explanation for the correlation between
$\Delta K\left(n \right)$ and $N\left(\varepsilon_F \right)$.
More intuitively, 
$\Delta K$ scales to the population of electrons around the fermi
surface [$\propto N\left(\varepsilon_F\right)$] which is subject to
the phonon scattering.
\par
\begin{figure}[p]
\begin{center}
\vspace{0mm}
\hspace{0mm}
\epsfxsize=17cm
\epsfbox{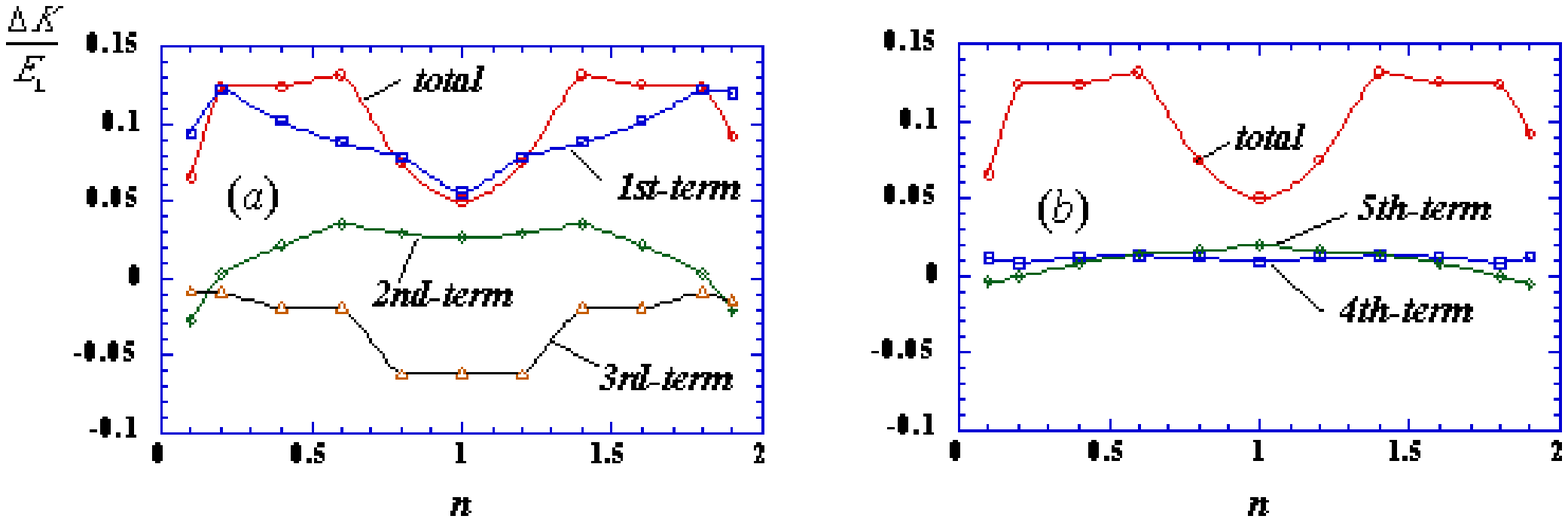}
\vspace{0mm}
\caption[aaa]{Filling ($n$-) dependence of the each contribution
defined in Eq. (\ref{Eq.J14}).}
\label{JF4}
\end{center}
\end{figure}

%
In order to see how each scattering process contributes, we re-divide
$\Delta K$ into several contributions as
%
\begin{eqnarray}
\label{Eq.J14}
\Delta K 
&=& \frac{1}{2}\sum\limits_k {\left[ {\varepsilon_k^{aa}\cdot I_{\left({aa;aa} 
\right)}^{\left({aa}\right)} + \varepsilon _k^{bb}  \cdot I_{\left( {bb;bb} 
\right)}^{\left( {bb} \right)} } \right]}  + \frac{1}{2}\sum\limits_k {\left[ 
{\varepsilon _k^{aa}  \cdot I_{\left( {ab;ab} \right)}^{\left( {aa} \right)}  
+ \varepsilon _k^{bb}  \cdot I_{\left( {ab;ab} \right)}^{\left( {bb} \right)} 
} \right]}
\nonumber \\
& &
 + \sum\limits_k {\left[ {\varepsilon _k^{ab}  \cdot \left( {I_{\left({ab;aa}
\right)}^{\left( {aa} \right)}  + I_{\left( {ab;bb} \right)}^{\left( {aa} 
\right)} } \right)} \right]}  + \sum\limits_k {\left[ {\varepsilon _k^{aa} 
\cdot I_{\left( {ab;aa} \right)}^{\left( {ab} \right)}  + \varepsilon_k^{bb}
\cdot I_{\left( {ab;bb} \right)}^{\left( {ab} \right)} } \right]} 
\nonumber \\
& &
 + \sum\limits_k {\left[ {\varepsilon _k^{ab}  \cdot \left( {I_{\left({ab;ab}
 \right)}^{\left( {ab} \right)}  + I_{\left( {aa;bb} \right)}^{\left( {ab} 
\right)} } \right)} \right]} \ ,
\end{eqnarray}
%
and plotted each contribution separately in Fig. \ref{JF4}.
%
\noindent
%
The first, second and  third terms are coming from $\Sigma_k^{\gamma\gamma}$, 
whereas the fourth and fifth from $\Sigma_k^{\gamma\bar\gamma}$.
The latter contribution is small compared with the former.
The first term, which is diagonal with respect to the suffices of $I$, 
just corresponds to the superposition of two breathing type diagrams with
$u$- and $v$-vertices, respectively.
(Remember the suffix rule of 
$I_{\left( {\gamma _2 \gamma '_2 ;\gamma _3 \gamma '_3 } \right)}
^{\left( {\gamma _1 \gamma '_1 } \right)}$ mentioned before).
The other terms arise due to the multiband structure and the
JT interaction, which are sorted further into two classes.
One class (the second and fourth terms ) corresponds to the 
twice inversion of the
orbital state (correspondingly the off-diagonal orbital suffix $ab$
appears twice) to come back to the  original orbital state
(like $a\rightarrow b\rightarrow a$).
Consequently this class picks up the diagonal dispersion $\varepsilon
_k^{\gamma\gamma}$ as its weight.
The other class (the third and the fifth terms) with odd number 
the orbital index $ab$ thus picks up the off-diagonal weight 
$\varepsilon_k^{ab}$ (like $a\rightarrow a\rightarrow b$).
Because $\varepsilon_k^{ab}$ roughly corresponds to the energy scale of
the stabilization due to the band hybridization, it gives basically
the negative contribution (stabilization of the energy) as shown
by the behavior of the third term in Fig. \ref{JF4}.
This stabilization around $n=1$ is mainly attributed to the dip with 
negative values of the "3rd-term" seen in Fig. \ref{JF4} (a).
This behavior is understood as follows.
The third term $\propto\varepsilon_k^{ab}$ reflects the 
stabilization of the lower band due to the
repulsion with the upper band
(its magnitude is $t_{i,i+\delta}^{ab}$).
Such a stabilization is most remarkable at the region where the hybridizing 
two bands cross with each other.
For the half-filled case, $n=1$, the fermi level is located at the middle
of the bandwidth, where the band-crossing occurs, leading to the most
effective stabilization.
That is why the stabilization of the third term is most remarkable
around $n=1$.
%
\section{Strong correlation limit}
%
Strong on-site repulsions in $e_g$ orbitals can be written as
\cite{MAE98a}
%
\begin{equation}
\label{Eq.J15}
H_{\rm on-site}=- \tilde \beta \sum\limits_j {\vec T_j }  \cdot \vec T_j
\quad\ ,\quad
\vec T_j  = \frac{1}{2}\sum\limits_{\gamma \gamma '} {d_{j\gamma }^\dag  
\vec \sigma _{\gamma \gamma '} d_{j\gamma '}} \ ,
\end{equation}
%
with spinless operators.
$\vec T_j$ is the isospin operator representing the orbital degrees of
freedom with $2 \times 2$ Pauli matrices 
$\vec \sigma _{\gamma\gamma'} = \left(\sigma^x_{\gamma\gamma'}
, \sigma^y_{\gamma\gamma'}, \sigma^z_{\gamma\gamma'}\right)$.
$\tilde\beta$ is a parameter of the electron-electron
interaction of the order of the Hubbard repulsive $U$.
This interaction induces the finite orbital polarization, which 
can be represented by the Stratonovich-Hubbard field 
(orbital fluctuation field) ${\vec \varphi}_T$ as\cite{MAE98a}
%
\begin{eqnarray}
\label{Eq.J16}
H_{el} &=& \sum\limits_{j,\gamma}{
d_{j\gamma}^\dag \left(\partial_\tau-\mu\right)d_{j\gamma}} +
\sum\limits_{i\delta ,\gamma \gamma '} {t_{i,i + \delta }^{\gamma 
\gamma '} d_{i\gamma }^\dag  d_{i + \delta ,\gamma '} }  + \sum\limits_j 
{\left[ {\frac{1}{{2M}}\vec P_j  \cdot \vec P_j  + \frac{{M\Omega ^2 }}{2}
\vec Q_j  \cdot \vec Q_j  + \tilde \beta \vec \varphi _T^2 } \right]}  
\nonumber \\ & &
- \sum\limits_j {\vec T_j  \cdot \left( {2\tilde \beta \vec \varphi _T 
 - g\vec Q_j } \right)} \ .
\end{eqnarray}
%
It is seem that the orbital fluctuation field ${\vec \varphi}_T$ 
as well as the JT phonon ${\vec Q}$ is coupled to the isospin
${\vec T}$ in the form of linear combination
$2\tilde\beta\vec\psi=2\tilde\beta\vec\varphi_T-g\vec Q_j$\cite{NAG98}.
After integrating out phonon coordinates, the effective
action in terms of the field $\psi$ is obtained as
%
\begin{eqnarray}
\label{Eq.J17}
S_{\rm{eff}} 
&=& \int_0^\beta {d\tau} 
\left[
\sum\limits_{j,\gamma}{
d_{j\gamma}^\dag \left(\partial_\tau-\mu\right)d_{j\gamma}}
+\sum\limits_{i\delta ,\gamma \gamma '} 
{t_{i,i + \delta }^{\gamma \gamma'} d_{i\gamma}^\dag d_{i+\delta ,
\gamma '} }  
\right]
+ \tilde \beta \sum\limits_{j,n} {\frac{{2\tilde \beta M
\left( {\omega _n^2  + \Omega ^2 } \right)}}{{2\tilde \beta M
\left( {\omega _n^2  + \Omega ^2 } \right) + g^2 }} \cdot } \vec 
\psi_{j,n}^*\psi_{j,n}  
\nonumber \\ &&
- 2\tilde \beta \sum\limits_{j,n} {\vec T_{j,n}  
\cdot \vec \psi _{j,n} } \ .
\end{eqnarray}
%
where $\omega_n = 2 \pi n/\beta $ is the Matsubara frequency
for the bosons.
The phonon dynamics induces the retardation effect for the field 
$\psi$, which is represented by 
the $\omega_n$-dependence of the second term in the above
equation.
Now let us assume that the electron correlation is much larger than 
the JT coupling, namely ${\tilde \beta} \gg E_L$.
It is noted here that we do not assume $E_L\ll t_0$, namely the
weak coupling limit.
Then we can expand in the JT coupling $g$ in eq.(17) as
%
\begin{eqnarray}
\label{Eq.J19}
S_{\rm eff}
&=& \int {d\tau} 
\left[
\sum\limits_{j,\gamma}{
d_{j\gamma}^\dag \left(\partial_\tau-\mu\right)d_{j\gamma}}
+\sum\limits_{i\delta ,\gamma \gamma '} 
{t_{i,i + \delta }^{\gamma \gamma '} d_{i\gamma }^\dag  d_{i+\delta ,
\gamma '} }  
\right]
\nonumber \\ &&
+ \tilde \beta \sum\limits_{j,n} {\left[ {1 - \frac{{g^2 }}{{2\tilde 
\beta M\left( {\omega _n^2  + \Omega ^2 } \right)}}} \right] \cdot } \vec 
\psi _{j,n}^* \vec \psi _{j,n} - 2\tilde \beta \sum\limits_{j,n} 
{\vec T_{j,n}  \cdot \vec \psi _{j,n} } \ .
\end{eqnarray}
%
Because we are now interested in the strong correlation limit,
${\tilde \beta}\gg t_0$, the magnitude of the orbital 
polarization is fully developed. This corresponds to the fixed 
$|{\vec  \psi}| = \varphi_T = 1/2$, and we consider its direction only within 
the $xz$-plane because the JT coupling prefers 
the real orbital states\cite{MAE98a}. Then ${\vec  \psi}$ is parametrized as 
%
\begin{equation}
\label{Eq.J22}
\vec \psi_j= \psi\cdot \ ^t \left(
\sin \theta _j, 0, \cos \theta _j 
\right) \ ,
\end{equation}
with the phase angle $\theta_j$ being the only relevant degrees of freedom.
Correspondingly, the isospin is forced to be parallel to 
$\vec \psi$, and hence the Grassman variables 
$d^\dagger_{j \gamma}$,$d_{j \gamma}$ are replaced by
%
\begin{eqnarray}
\label{Eq.vec}
d_{j\gamma} &=& [ f_j \cos(\theta_j/2), f_j \sin(\theta_j/2) ] 
\nonumber \\
d^\dagger_{j\gamma} &=& ^t [ f^\dagger_j \cos(\theta_j/2), 
f^\dagger_j \sin(\theta_j/2) ] \ , 
\end{eqnarray}
with the spin/orbital-less fermion variable $f^\dagger, f$.
Putting this expression into Eq. (\ref{Eq.J19}), the kinetic energy term
can be written as
%
\begin{equation}
\label{Eq.J22}
\sum_{i, \delta} t_{i, i+\delta}(\theta_i, \theta_{i + \delta} ) 
\cdot f^\dagger_i f_{i + \delta} \ ,
\end{equation} 
%
with the $\theta$-dependent transfer integral,
%
\begin{eqnarray}
t_{i, i+\delta}(\theta_i, \theta_{i + \delta} )
&=&
t_{i , i+\delta}^{11} \cos(\theta_i/2) \cos(\theta_{i+\delta}/2) +
t_{i , i+\delta}^{22} \sin(\theta_i/2) \sin(\theta_{i+\delta}/2)
\nonumber \\
&+& t_{i , i+\delta}^{12} \cos(\theta_i/2) \sin(\theta_{i+\delta}/2) +
t_{i , i+\delta}^{21} \sin(\theta_i/2) \cos(\theta_{i+\delta}/2) \ .
\end{eqnarray}
%
This gives the coupling of $\theta$-field to the fermion.
On the other hand, the dynamics of the $\theta$-field is generated through this
coupling by integrating over the fermions $f^\dagger, f$.
%
\begin{equation}
\label{Eq.K25}
S_{\rm eff}^0 = \sum_{q, \omega_n} 
\Pi(\vec q, \omega_n)\cdot\theta(\vec q,\omega_n) \theta(-\vec q,-\omega_n) \ ,
\end{equation}
%
where $\theta$ is measured from the mean field value, and
$\Pi(\vec q,\omega_n)$ is the orbital correlation function of 
the fermions. 
Although the quantitative results depend on the details of the model 
and the orbital ordering to start with,
the orbital fluctuation has a gap of the order of $t_0$
and there occurs no infrared divergence. 
Therefore the characteristic frequency $\omega_n$ and wavevector $\vec q$ for 
$\Pi(\vec q,\omega_n)$ are $\sim t_0$, and $\pi/a$
($a$: lattice constant), respectively.
The Hamiltonian consists only of the kinetic energy besides
the JT coupling term, and its expectation value is given by
%
\begin{equation}
\label{Eq.energy}
\left\langle H_0 \right\rangle = 
\sum_{i, \delta} t_{i, i+\delta}(\theta_i=0, \theta_{i + \delta}=0 ) 
\left\langle f^\dagger_i f_{i + \delta}\right\rangle_{\rm mean \   field}
+ \lim_{\beta \to \infty} 
{ 1 \over {2 \beta}} \sum_{{\vec q},\omega} 
\ln \Pi({\vec q},\omega) \ .
\end{equation}
%
Now let us analyse the correction term due to the JT interaction 
%
\begin{equation}
\label{Eq.J21}
\delta S_{\rm eff}  =  - \sum\limits_{j,n} 
{\frac{{g^2 }}{{2M\left( {\omega _n^2  + \Omega ^2 } \right)}} \cdot } 
\vec \psi _{j,n}^* \vec \psi _{j,n}
= \sum\limits_{j,n} {A\left( {i\omega _n } \right) \cdot } \vec \psi _{j,n}^* \vec \psi _{j,n} \ .
\end{equation}
%
The differential operator $A\left(i\omega_n\right)$ on $\vec\psi_{j,n}$ 
leads to the dynamics of the phase angle $\partial_\tau\theta_j$ through 
the component,
%
\begin{equation}
\label{Eq.J23}
\left( {\partial _\tau  \vec \psi _j} \right)^*  \cdot \left( {\partial _\tau  
\vec \psi_j} \right)= \psi _j^2  \cdot \left( {\partial _\tau  \theta _j } 
\right)^2 \ .
\end{equation}
%
Hence the contribution to the dynamics of $\theta$-field due to
JT coupling is given by
%
\begin{equation}
\label{Eq.J24}
\delta S_{\rm eff}  
= \sum\limits_{j,n} {A\left( {i\omega_n } \right)\cdot} 
\vec \psi_{j,n}^* \vec\psi _{j,n}  \to \psi^2 \cdot\sum\limits_{j,n} 
{[
A\left(i\omega_n\right)-A\left(0\right) 
]} \cdot
\theta_{j,n} \theta_{j ,- n} \ .
\end{equation}
%
Now the dymanics of the orbital is determined by the propagator
$D({\vec q},\omega_n)$ defined by
%
\begin{equation}
\label{Eq.prop}
[D({\vec q},\omega_n)]^{-1} 
= \Pi({\vec q},\omega_n) + \frac{\psi^2}{2} \cdot E_L 
{ {\omega_n^2} \over { \omega_n^2 + \Omega^2} }.
\end{equation}
%
There are two limits of interest. In the case of weak coupling, i.e.,
$E_L\ll t_0$, the dynamics of the orbital is determined by 
$\Pi(\vec q,\omega_n)$ and the characteristic energy is
of the order of $t_0$. Therefore we can replace $\omega_n$ in 
Eq. (\ref{Eq.prop}) by $\sim t_0 \gg\Omega$, and the 
correction of the propagator is of the order of $E_L/t_0^2$.
More explicitly the reduction of the kinetic energy gain $\Delta K$
due to JT coupling is estimated by replacing $\ln \Pi$ in
Eq. (\ref{Eq.energy}) by $-\ln D$ in Eq. (\ref{Eq.prop}) as
%
\begin{equation}
\label{Eq.deltaK}
\Delta K \sim \int d \omega \cdot
{ {E_L \omega^2/(\omega^2 + \Omega^2)} \over { \Pi(\vec q,\omega) }}
\sim E_L \omega_c^2/(\omega_c^2 + \Omega^2) \ ,
\end{equation}
%
where $\omega_c$ is the characteristic frequency of the orbital fluctuation
and $\omega_c \sim t_0$.
This energy correction quadratically grows up with increasing 
$\omega_c/\Omega \ll 1$ and then saturates into the lattice relaxation
energy $E_L=g^2/M\Omega^2$ with $\omega_c/\Omega \gg 1$.
Considering that $\omega_c \sim t_0 \gg \Omega$, we conclude that 
$\Delta K \sim E_L$. As increasing $E_L (\sim t_0 )$, we expect the
saturation effect as $\Delta K \sim E_L t_0/(t_0 + E_L)$ as is evident from 
eq.(28).

The strong couling limit, i.e., $E_L\gg t_0$, is more
interesting. In this case, the orbital dynamics is determined by both 
$\Pi$ and $E_L$ terms in Eq. (\ref{Eq.prop}). 
More explicitly the propagator is approximated as
%
\begin{equation}
\label{Eq.prop1}
[D({\vec q},\omega_n)]^{-1} 
\cong  t_0 + \frac{\psi^2}{2} \cdot E_L 
{ {\omega_n^2} \over { \omega_n^2 + \Omega^2} },
\end{equation}
%
and its characteristic energy scale is then given by
%
\begin{equation}
\label{Eq.omegac}
\omega_c \cong \Omega 
\sqrt{ { {t_0} \over { E_L} } }\ll \Omega.
\end{equation}
%
Therefore there occurs the slow down of the orbital motion in this 
strong coupling limit. However the small polaron effect should be relevant 
in this case, which can not be treated in the present formalism, and the
detailed study on this strong coupling case is left for future studies.
%
\section{discussions}
%
\par
%
We now compare the results for the non-interacting and strongly interacting
limits. 
The order estimation of $\Delta K_{U=0}$ for the non-interacting limit
is as follows.
Rather complicated form of the non-interacting result, Eq. (\ref{Eq.J13}), 
Eq. (\ref{Eq.JA3}), Eq. (\ref{Eq.JA6}), and Eq. (\ref{Eq.JA7}), roughly
takes the  $\Omega$-dependence as
%
\begin{equation}
\label{Eq.J27}
\Delta K_{U=0}\sim 
t_0\!\cdot\!\left(\frac{g}{\sqrt{M\Omega}}\right)^2
\!\cdot\!\frac{1}{\left(t_0+\Omega\right)^2} \ .
=
t_0\cdot\frac{g^2}{M\Omega}\!\cdot\!\frac{1}{\left(t_0+
\Omega\right)^2}
=
E_L  \cdot  { { \Omega/t_0} \over {(1 + \Omega/t_0)^2} }
\end{equation}
%
The dependence is hence a kind of perturbative forms
with the intermediate energy denominator 
$1/\left(t_0+\Omega\right)^2$ and the vertex  $g/\sqrt{M\Omega}$.
The small factor $\Omega/t_0$ comes from the fact that only the 
states with the energy window $\sim \Omega$ near the Fermi energy
is influenced by the el-ph interaction. 
More explicitely, 
\par
\begin{equation}
\label{Eq.J29}
\frac{\Delta K}{K_0} = \frac{\Delta K/E_L}{K_0/E_L}
=\frac{\left(\rm value\ picked\ up\ from\ Fig. 4\right)}{K_0/E_L} \ .
\end{equation}
%
$E_L$ can be evaluated as $\sim 0.6$ eV from the literature 
\cite{MIL96c}.
With $K_0 \sim 2.16$ eV in our caluculation, and the value $\sim 0.1$
in Fig. \ref{JF4}, we get $\Delta K/K_0 \sim 3 \%$ as a lower bound.
\par
On the other hand, in the strong correlation limit,
the effect of Fermi degeneracy and the small factor $\Omega/t_0$ are
missing and $\Delta K \sim E_L$.
This means that the strong correlation enhances the JT effect.
%
%
It is reported that the observed spin stiffness is well reproduced
even semiquantitatively by meanfield estimations without considering
the el-ph interaction discussed here. \cite{MAE99a,QUI98}
This means that the small polaronic effect is absent in the 
metallic region, namely $E_L$ is of the order or less than 
$t_0$ (Here $t_0$ should be interpreted as the ``bandwidth'' rather
than the transfer integral). It is reasonable with the above estimation 
that $E_L \sim 0.6$ eV. 
\par
%
In summary, we have studied JT el-ph  effect in three dimension with
and without the electron correlations.
In the non-interacting limit, the reduction is calculated as a function 
of the doping concentration.
In this case, the doping dependence is mainly dominated by the density of
states at the  Fermi energy.
It is shown that the kinetic energy is always reduced 
by the JT el-ph interaction even if the off-diagonal processes in orbital 
indices are taken into account. The reduction $\Delta K$ of the kinetic energy
$K$ is estimated as $\Delta K/K \cong 3 \%$ in this non-interacting case.
This small value is due to the small factor $\Omega/t_0$ occuring 
in the Fermi degenerate case.
In the strong correlation limit, we have derived  an
effective action to study the el-ph interaction.
The small factor $\Omega/t_0$ is missing in this case and
el-ph interaction is enhanced by the strong correlation.
$\Delta K \sim E_L t_0/(E_L + t_0)$ is this case without the 
small polaron formation.
From the comparison with experiments and the above results,
the small polaron formation is unlikely in the metallic state.
\par
%
The authors would like to thank A. Millis and H. Yoshizawa for their 
valuable discussions.
This work was supported by Priority Areas Grants from the Ministry
of Education, Science and Culture of Japan.
R.M. is supported by the Japan Society for the Promotion of Science (JSPS) 
for Young Scinentists during this work.
%
\appendix
%
\section{evaluation of Eq. (12)}
\begin{figure}[p]
\begin{center}
\vspace{0mm}
\hspace{0mm}
\epsfxsize=10cm
\epsfbox{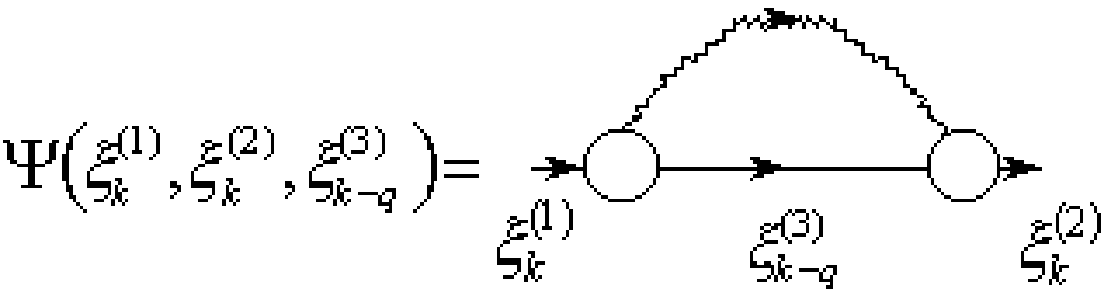}
\vspace{0mm}
\caption[aaa]{Diagram corresponding to 
$\Psi \left( {\xi _k^{\left( 1 \right)} ,\xi _k^{\left( 2 \right)} ,
\xi _{k - q}^{\left( 3 \right)} } \right)$}
\label{JF5}
\end{center}
\end{figure}

%
With $\sigma_k\left(z,\xi_{k-q}\right)$ defined by the breathing
type self-energy expression \cite{MAH90} as,
%
\begin{equation}
\label{Eq.JA1}
\Sigma _{\vec k}^0\left( z \right)={{g^2} \over {2N\beta M\omega }}
\sum\limits_q {\left[ {{{f\left( {\xi _{k-q}^{}} \right)+N\left( \omega  
\right)} \over {z-\xi _{k-q}^{-\omega }}}-{{f\left( {\xi _{k-q}^{}} \right)
+N\left( {-\omega } \right)} \over {z-\xi _{k-q}^{+\omega }}}} \right]}
\equiv\sum\limits_q {\sigma _k\left( {z,\xi _{k-q}^{}} \right)}\ , 
\end{equation}
%
the JT type self-energy Eq. (11) can be written as
%
\begin{equation}
\label{Eq.JA2}
\Sigma_{\vec k}^{\gamma\gamma'}\left( z \right)\!=\!\sum\limits_q {\left[ 
{\left({A_{+;k-q}^{\gamma\gamma'}+A_{+;k-q}^{\bar\gamma\bar\gamma'}} 
\right)\!\cdot\!\sigma_k\left( {z,\Xi_{k-q}^{\left( \!+\!\right)}} \right)
+\left( {A_{-;k-q}^{\gamma\gamma'}+A_{-;k-q}^{\bar \gamma \bar \gamma'}} 
\right)\!\cdot\!\sigma _k\left( {z,\Xi _{k-q}^{\left( - \right)}} \right)} 
\right]}\ ,
\end{equation}
%
with coefficients $A_{\pm;k}^{\gamma\gamma'}$ defined in Eq. (\ref{Eq.J8}).
$N\left(\omega\right)$ represents the bose distribution function.
The notation $\xi_k^{\left(\pm;\Omega\right)}$ is defined as 
$\xi_k^{\left(\pm;\Omega\right)}=\xi_k^{\left(\pm\right)}+\Omega$.
Substituting Eq. (\ref{Eq.JA2}) into Eq. (\ref{Eq.J12}) leads to
%
\begin{eqnarray}
\label{Eq.JA3}
\frac{
I_{\left({\gamma_2\gamma'_2;\gamma_3\gamma'_3}\right)}
^{\left({\gamma_1\gamma'_1}\right)}
}{g^2/2NM\omega}
&=&
E_{k-q}^{\gamma_1\gamma'_1\left(+\right)}A_{+;k}
^{\gamma_2\gamma'_2}A_{+;k}^{\gamma_3\gamma'_3}\cdot
\Psi\left({\Xi_k^{\left(+\right)},\Xi_k^{\left(+\right)},
\Xi_{k-q}^{\left(+\right)}}\right)
\nonumber\\
&&
+E_{k-q}^{\gamma_1\gamma'_1\left(+\right)}A_{-;k}^{\gamma_2\gamma'_2}A_{-;k}^{\gamma_3\gamma'_3}\cdot\Psi\left({\Xi_k^{\left(-\right)},\Xi_k^{\left(-\right)},\Xi_{k-q}^{\left(+\right)}}\right)
\nonumber\\
&&
+\left({E_{k-q}^{\gamma_1\gamma'_1\left(+\right)}A_{+;k}^{\gamma_2\gamma'_2}A_{-;k}^{\gamma_3\gamma'_3}+E_{k-q}^{\gamma_1\gamma'_1\left(+\right)}A_{-;k}^{\gamma_2\gamma'_2}A_{+;k}^{\gamma_3\gamma'_3}}\right)\cdot\Psi\left({\Xi_k^{\left(+\right)},\Xi_k^{\left(-\right)},\Xi_{k-q}^{\left(+\right)}}\right)
\nonumber\\
&&
+E_{k-q}^{\gamma_1\gamma'_1\left(-\right)}A_{+;k}^{\gamma_2\gamma'_2}A_{+;k}^{\gamma_3\gamma'_3}\cdot\Psi\left({\Xi_k^{\left(+\right)},\Xi_k^{\left(+\right)},\Xi_{k-q}^{\left(-\right)}}\right)
\nonumber\\&&
+E_{k-q}^{\gamma_1\gamma'_1\left(-\right)}A_{-;k}^{\gamma_2\gamma'_2}A_{-;k}^{\gamma_3\gamma'_3}\cdot\Psi\left({\Xi_k^{\left(-\right)},\Xi_k^{\left(-\right)},\Xi_{k-q}^{\left(-\right)}}\right)
\nonumber\\&&
+\left({E_{k-q}^{\gamma_1\gamma'_1\left(-\right)}A_{+;k}^{\gamma_2\gamma'_2}A_{-;k}^{\gamma_3\gamma'_3}+E_{k-q}^{\gamma_1\gamma'_1\left(-\right)}A_{-;k}^{\gamma_2\gamma'_2}A_{+;k}^{\gamma_3\gamma'_3}}\right)\cdot\Psi\left({\Xi_k^{\left(+\right)},\Xi_k^{\left(-\right)},\Xi_{k-q}^{\left(-\right)}}\right) \ ,
\end{eqnarray}
%
where we defined
%
\begin{equation}
\label{Eq.JA4}
E_{k-q}^{\gamma_1\gamma'_1\left(\pm\right)}=A_{\pm;k-q}^{\gamma_1\gamma'_1}+A_{\pm;k-q}^{\bar\gamma_1\bar\gamma'_1} \ ,
\end{equation}
[the first (second) term corresponds to the scattering by $u$- ($v$-) 
phonon, respectively, as in Fig. 2].
Function $\Psi$ is defined as the integral
\begin{equation}
\label{Eq.A5}
\Psi \left( {\xi _k^{\left( 1 \right)} ,\xi _k^{\left( 2 \right)} ,\xi _{k - q}^{\left( 3 \right)} } \right)
=
\oint_c {\frac{dz }{2\pi i} \cdot f\left( z \right)}\cdot \frac{{\sigma _k \left( {z,\xi _{k - q}^{\left( 3 \right)} } \right)}}{{\left( {z - \xi _k^{\left( 1 \right)} } \right)\left( {z - \xi _k^{\left( 2 \right)} } \right)}}\ ,
\end{equation}
%
corresponding to a diagram shown in Fig. \ref{JF5}.
%
\noindent
%
Depending on the degree of the pole, it is evaluated as,
%
\begin{eqnarray}
\label{Eq.JA6}
\Psi\left({\xi_k^{\left( 1 \right)} ,\xi _k^{\left( 2 \right)} ,
\xi _{k - q}^{\left( 3 \right)} } \right)
&=&
f\left( {\xi _{k - q}^{\left( 3 \right)} } \right) \cdot 
\left[
{\frac{{f\left( {\xi _k^{\left( 1 \right)} } \right)}}{{\left( {\xi _k^{\left( 1 \right)}  - \xi _{k - q}^{\left( {3; - \omega } \right)} } \right)\left( {\xi _k^{\left( 1 \right)}  - \xi _k^{\left( 2 \right)} } \right)}} + \frac{{f\left( {\xi _k^{\left( 2 \right)} } \right)}}{{\left( {\xi _k^{\left( 2 \right)}  - \xi _k^{\left( 1 \right)} } \right)\left( {\xi _k^{\left( 2 \right)}  - \xi _{k - q}^{\left( {3; - \omega } \right)} } \right)}}}
\right.
\nonumber\\
& &\left.\quad\quad\quad\quad\quad\quad\quad\quad\quad
{ + \frac{{f\left( {\xi _{k - q}^{\left( {3; - \omega } \right)} } \right)}}{{\left( {\xi _{k - q}^{\left( {3; - \omega } \right)}  - \xi _k^{\left( 1 \right)} } \right)\left( {\xi _{k - q}^{\left( {3; - \omega } \right)}  - \xi _k^{\left( 2 \right)} } \right)}}}\right]
\nonumber \\
& & + \bar f\left( {\xi _{k - q}^{\left( 3 \right)} } \right) \cdot 
\left[
{\frac{{f\left( {\xi _k^{\left( 1 \right)} } \right)}}{{\left( {\xi _k^{\left( 1 \right)}  - \xi _{k - q}^{\left( {3; + \omega } \right)} } \right)\left( {\xi _k^{\left( 1 \right)}  - \xi _k^{\left( 2 \right)} } \right)}} + \frac{{f\left( {\xi _k^{\left( 2 \right)} } \right)}}{{\left( {\xi _k^{\left( 2 \right)}  - \xi _k^{\left( 1 \right)} } \right)\left( {\xi _k^{\left( 2 \right)}  - \xi _{k - q}^{\left( {3; + \omega } \right)} } \right)}}}
\right.
\nonumber \\
& &\left.\quad\quad\quad\quad\quad\quad\quad\quad\quad
{ + \frac{{f\left( {\xi _{k - q}^{\left( {3; + \omega } \right)} } \right)}}{{\left( {\xi _{k - q}^{\left( {3; + \omega } \right)}  - \xi _k^{\left( 1 \right)} } \right)\left( {\xi _{k - q}^{\left( {3; + \omega } \right)}  - \xi _k^{\left( 2 \right)} } \right)}}}
\right] \ ,
\end{eqnarray}
%
for $\xi _k^{\left( 1 \right)}\ne\xi_k^{\left( 2 \right)}$ and
%
\begin{eqnarray}
\label{Eq.JA7}
\Psi \left( {\xi _k,\xi _k,\xi _{k - q}} \right)
&=&
f{\left( {\xi _{k - q}^{} } \right)\frac{{f'\left( {\xi _k^{} } \right)\left( {\xi _k^{}  - \xi _{k - q}^{ + \omega } } \right) - f\left( {\xi _k^{} } \right) + f\left( {\xi _{k - q}^{ + \omega } } \right)}}{{\left( {\xi _k^{}  - \xi _{k - q}^{ + \omega } } \right)^2 }}}
\nonumber \\
& &\quad\quad\quad\quad\quad\quad + \bar f\left( {\xi _{k - q}^{} } \right)\frac{{f'\left( {\xi _k^{} } \right)\left( {\xi _k^{}  - \xi _{k - q}^{ - \omega } } \right) - f\left( {\xi _k^{} } \right) + f\left( {\xi _{k - q}^{ - \omega } } \right)}}{{\left( {\xi _k^{}  - \xi _{k - q}^{ - \omega } } \right)^2 }}
\nonumber \\
&=&- f\left( {\xi _k} \right) \left[
{f\left( {\xi _{k - q}^{} } \right)\bar f\left( {\xi _{k - q}^{ - \omega } } \right) \cdot P\left[ {\frac{1}{{\left( {\xi _k^{}  - \xi _{k - q}^{ - \omega } } \right)^2 }}} \right]}
\right.
\nonumber \\
& &\quad\quad\quad\quad\quad\quad\quad\quad\quad \left.
{ + \bar f\left( {\xi _{k - q}^{} } \right)\bar f\left( {\xi _{k - q}^{ + \omega } } \right) \cdot P\left[ {\frac{1}{{\left( {\xi _k^{}  - \xi _{k - q}^{ + \omega } } \right)^2 }}} \right]}\right]
\nonumber \\
& &\quad + \bar f\left( {\xi _k} \right)\left[
f\left( {\xi _{k - q}^{} } \right)f\left( {\xi _{k - q}^{ - \omega } } \right) \cdot P\left[ {\frac{1}{{\left( {\xi _k^{}  - \xi _{k - q}^{ - \omega } } \right)^2 }}} \right]
\right.
\nonumber \\ & &\quad\quad\quad\quad\quad\quad\quad\quad\quad
 + \bar f\left( {\xi _{k - q}^{} } \right)\left( {\xi _{k - q}^{ + \omega } } \right) \cdot P\left[ {\frac{1}{{\left( {\xi _k^{}  - \xi _{k - q}^{ + \omega } } \right)^2 }}} \right] 
\nonumber \\ &\equiv &
- \left( {\Psi _{k,q}^{\left( {1,2} \right)}  + \Psi _{k,q}^{\left( {3,4} \right)} } \right) \ ,
\end{eqnarray}
%
where $\bar f\left({\xi_k}\right)=1-f\left({\xi_k}\right)$.
With notations here, the expression for the simple breathing type case
with the single band system \cite{KIM99} is expressed as
%
\begin{equation}
\label{Eq.JA8}
\Delta K^{\rm single}={{g^2} \over {4NM\omega }}\sum\limits_{kq} 
{\left[ {-\varepsilon_k\cdot \Psi \left( {\xi_k,\xi_k,\xi _{k-q}} 
\right)} \right]} \ .
\end{equation}
%
\section{Property of the function $\Psi$ in Eq. (A7)}
\begin{figure}[p]
\begin{center}
\vspace{0mm}
\hspace{0mm}
\epsfxsize=17cm
\epsfbox{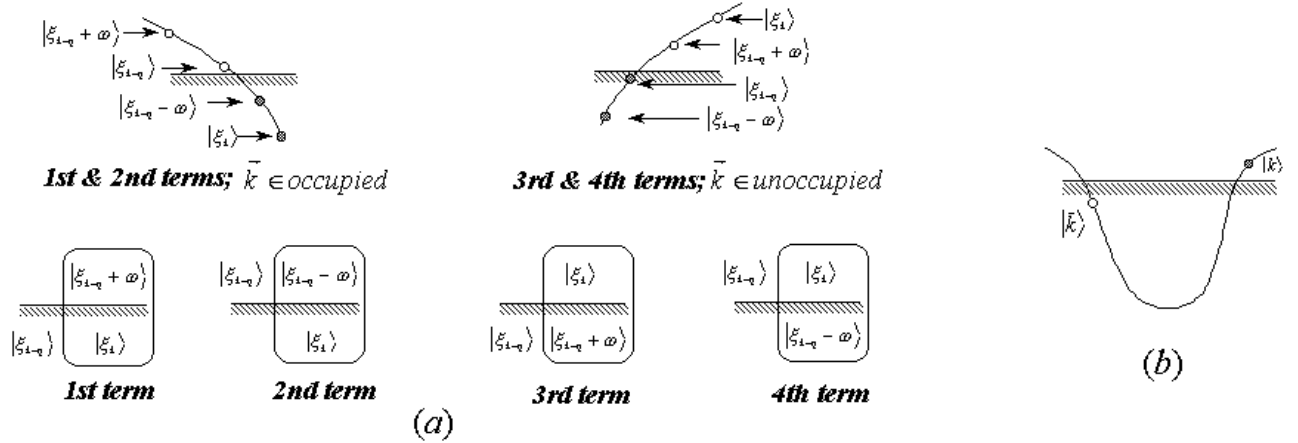}
\vspace{0mm}
\caption[aaa]{Configurations of $\left|\xi_k\right\rangle$, 
$\left|\xi_{k-q}\right\rangle$, and $\left|\xi_{k-q}\pm\omega\right\rangle$
under which each term of Eq. (\ref{Eq.JA7}) contributes.}
\label{JFB1}
\end{center}
\end{figure}

\begin{figure}[p]
\begin{center}
\vspace{0mm}
\hspace{0mm}
\epsfxsize=17cm
\epsfbox{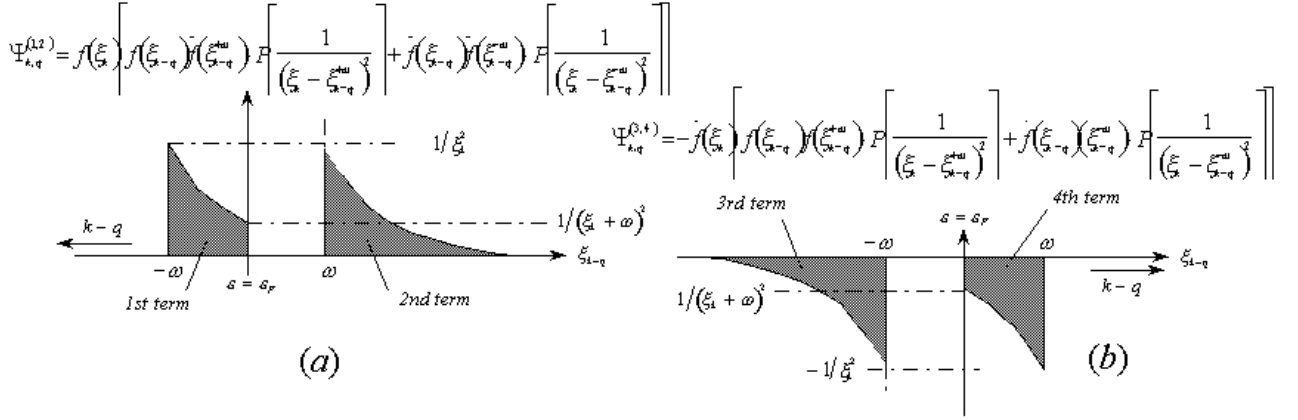}
\vspace{0mm}
\caption[aaa]{Behavior of each contribution of terms in Eq. (\ref{Eq.JA7}) 
as a function of $\xi_{k-q}$ for fixed $k$.}
\label{JFB2}
\end{center}
\end{figure}

%
For the simplest case with single-band electrons and 
breathing type phonons, Eq. (\ref{Eq.JA8}), the filling 
dependence is determined by 
$\sum\limits_{kq} 
{\left[ {-\varepsilon_k\cdot \Psi \left( {\xi_k,\xi_k,\xi _{k-q}} 
\right)} \right]}$.
Each term of the function $\Psi\left({\xi_k,\xi_k,\xi_{k-q}}\right)$
in Eq. (\ref{Eq.JA7}) contributes when the states
$\left|\xi_k\right\rangle$, $\left|\xi_{k-q}\right\rangle$, 
and $\left|\xi_{k-q}\pm\omega\right\rangle$
are in such configurations as shown in Fig. \ref{JFB1}(a).
%
\noindent
%
For given and fixed $k$, the contribution due to each term 
behaves as shown in Fig. \ref{JFB2} as a function of $\xi_{k-q}$.
Notice that the first and the second terms ($\Psi_{k,q}^{\left(1,2\right)}$)
are exclusive to the third and fourth terms
($\Psi_{k,q}^{\left(3,4\right)}$) each other.
Introducing a notation $\bar k \equiv k - 2k_F $ 
[see Fig. \ref{JFB1} (b)], Eq. (\ref{Eq.JA8}) is evaluated as,
%
\begin{eqnarray}
\label{Eq.JB1}
\Delta K^{\rm single}
&\sim&
\sum\limits_{kq}{\varepsilon_k\cdot\Psi_{k,q}^{\left({1,2}\right)}}
+\sum\limits_{kq}{\varepsilon_k\cdot\Psi_{k,q}^{\left({3,4}\right)}}
=\sum\limits_{k\in occupied}{\left({\varepsilon_k\cdot\sum\limits_q
{\Psi_{k,q}^{\left({1,2} \right)}}
+\varepsilon_{\bar k}\cdot\sum\limits_q{\Psi_{\bar k,q}^{\left({3.4}\right)
}}}\right)}  
\nonumber \\
&=&
\sum\limits_{k \in occupied}{\left({\varepsilon_k\cdot S_k^{\left({1,2} 
\right)}  
+ \varepsilon_{\bar k}\cdot S_{\bar k}^{\left( {3,4} \right)} } \right)} \ ,
\end{eqnarray}
%
where $S_k^{\left(1,2\right)}>0$ and $S_k^{\left(3,4\right)}<0$ are the
quadratures of the shaded areas with signs depicted in Fig. \ref{JFB2} 
(a) and (b), respectively.
%
\noindent
%
It can be written as
$S_{\bar k}^{\left(3,4\right)}=-S_k^{\left(1,2\right)}+ \delta S_k$
with a small deviation $\delta S_k$ which reflects the difference between
the band curvature at $\left|k\right\rangle$ and $\left|\bar k\right\rangle$
shown in Fig. \ref{JFB1} (b).
$\delta S_k$ vanishes when the system is half-filled where the fermi level
locates at the middle of the band.
Equation (\ref{Eq.JB1}) is then evaluated as, 
%
\begin{equation}
\label{Eq.JB2}
\Delta K^{single}\sim
\!\!\!\!\!\!
\sum\limits_{k\in occupied}{
\!\!\!\!\!\!
\left[{\left({\mu+\xi_k}
\right)S_k^{\left( {1,2} \right)}+ \left( {\mu-\xi_k }\right)\left({
-S_k^{\left( {1,2} \right)}+ \delta S_k } \right)} \right]}  
= 2
\!\!\!\!\!\!
\sum\limits_{k \in occupied} {
\!\!\!\!\!\!
\xi _k S_k^{\left(1,2 \right)} }  
+ \delta K \ ,
\end{equation}
%
with a small $\delta K$ compared with the first term.
Because $\xi_k\le 0$ for occupied states the result has therefore 
definite sign.
From Fig. \ref{JFB2}, it is understood that the contributions mainly
come from the vicinity of the fermi level.
It thus means that the result scales to the density of states at
the fermi level.
%

%
\end{document}